\documentclass[twocolumn]{aastex631}

\shorttitle{Catalog of over 10k SN-like ZTF light curves}
\shortauthors{Garretson et al.}

\graphicspath{{./}{figures/}}

\begin{document}

\title{Supernova host galaxy association and photometric classification of over 10,000 light curves\\ from the Zwicky Transient Facility }

\author[0000-0001-6922-8319]{Braden Garretson}
\affiliation{Department of Physics and Astronomy, Purdue University, 525 Northwestern Avenue, West Lafayette, IN 47907-2036, USA}

\author[0000-0002-0763-3885]{Dan Milisavljevic}
\affiliation{Department of Physics and Astronomy, Purdue University, 525 Northwestern Avenue, West Lafayette, IN 47907-2036, USA}
\affiliation{Integrative Data Science Initiative, Purdue University, West Lafayette, IN 47907, USA}

\author[0000-0002-1521-0479]{Jack Reynolds}
\affiliation{Department of Physics and Astronomy, Purdue University, 525 Northwestern Avenue, West Lafayette, IN 47907-2036, USA}

\author[0000-0002-8360-0831]{Kathryn E.\ Weil}
\affiliation{Department of Physics and Astronomy, Purdue University, 525 Northwestern Avenue, West Lafayette, IN 47907-2036, USA}

\author[0000-0001-8073-8731]{Bhagya Subrayan}
\affiliation{Department of Physics and Astronomy, Purdue University, 525 Northwestern Avenue, West Lafayette, IN 47907-2036, USA}

\author[0000-0003-0776-8859]{John Banovetz}
\affiliation{Department of Physics and Astronomy, Purdue University, 525 Northwestern Avenue, West Lafayette, IN 47907-2036, USA}

\author[0000-0002-7482-5078]{Rachel Lee}
\affiliation{Department of Physics and Astronomy, Purdue University, 525 Northwestern Avenue, West Lafayette, IN 47907-2036, USA}

\begin{abstract}

Here we present a catalog of 12,993 photometrically-classified supernova-like light curves from the Zwicky Transient Facility, along with candidate host galaxy associations. By training a random forest classifier on spectroscopically classified supernovae from the Bright Transient Survey, we achieve an accuracy of 80\% across four supernova classes resulting in a final data set of 8,208 Type Ia, 2,080 Type II, 1,985 Type Ib/c, and 720 SLSN. Our work represents a pathfinder effort to supply massive data sets of supernova light curves with value-added information that can be used to enable population-scale modeling of explosion parameters and investigate host galaxy environments. 

\end{abstract}

\keywords{\href{http://astrothesaurus.org/uat/1954}{Light curve classification(1954)} --- \href{http://astrothesaurus.org/uat/1668}{Supernovae(1668)} --- \href{http://astrothesaurus.org/uat/83}{Astronomy databases(83)}}

\section{Introduction} \label{sec:intro}

The upcoming Legacy Survey of Space and Time (LSST) by the Vera Rubin Observatory will provide exciting opportunities to investigate hundreds of thousands of supernovae and their host galaxy environments \citep{Ivezi__2019}.  However, this large influx of data also comes with numerous technical challenges, including the classification of supernovae, where techniques often suffer from limited and biased training data \citep{boone_avocado_2019,villar_superraenn_2020,sravan_real-time_2020,S_nchez_S_ez_2021}. Within this context, we were motivated to develop a supernova classifier for the alert stream of the Zwicky Transient Facility  \citep[ZTF;][]{Bellm_2018} and provide a catalog of supernova-like light curves for as many events as possible.

\section{Classification Pipeline} \label{sec:pipeline}
 
\subsection{Training Data} \label{subsec:training data}
Our classifier training data was built using supernovae observed by ZTF that have been spectroscopically classified by the Bright Transient Survey \citep[BTS;][]{Perley_2020} with the condition that the light curve contained alerts before and after the peak of the event. We split the data into 4 main types: Type Ia (2,247), Type II (534), Type Ib/c (179), and SLSN (39). Treatment of this class imbalance, which is amplified by our data being skewed towards bright, low redshift objects, is discussed in \S \ref{sec:imbalance}.

\subsection{Preprocessing with Gaussian Process Regression} \label{sec:gpr}
To mitigate challenges associated with sparse and incomplete data, we employed Gaussian processed regression (GPR) which allows us to use the available data and uncertainties in both bands to interpolate the missing data points. We fit a GPR using the python package George \citep{hodlr} following the method introduced by \cite{boone_avocado_2019}. We then correct each light curve for Milky Way extinction using \citet{1998ApJ...500..525S}.

\begin{deluxetable*}{lllllcllc}[!htp]
%\tablewidth{800pt}
\tablecolumns{9}
\tabletypesize{\scriptsize}
\tablehead{
\colhead{ZTF ID} & \colhead{SN Ia} \vspace{-0.35cm} & 
\colhead{SN II} & \colhead{SN Ib/c} & 
\colhead{SLSN} & \colhead{Host?} & \colhead{Host R.A.} & \colhead{Host Decl.} & \colhead{TNS} \\ \colhead{} & \colhead{Probability} & \colhead{Probability} & \colhead{Probability} & \colhead{Probability} & \colhead{} & \colhead{} & \colhead{} & \colhead{Classification}
} 

\startdata
ZTF20acoqlav & 0.513   &   0.097   &   0.118 & 0.271 & Galaxy & 1.7898 & 39.1709 & SN Ia \\ 
ZTF20acknpig & 0.022 & 0.857& 0.063& 0.056 & Galaxy & 30.7454 & 45.0238 & SN II \\ 
ZTF21aaehrnw & 0.103 & 0.248 & 0.058 & 0.590 & No Host & N/A & N/A & N/A \\
ZTF21aabyifm & 0.087& 0.258 &     0.637& 0.017 & Galaxy & 187.4084 & 9.4953 & SN Ib/c \\ 
ZTF21aapqbyz & 0.085 & 0.744 & 0.069    &  0.100 & Galaxy & 151.2126 & 48.0672 & N/A \\ 
ZTF20aczshxu & 0.477& 0.142& 0.191& 0.188 & Galaxy & 175.6201 &	54.2471 & N/A \\ 
ZTF20aawkgxa & 0.054     & 0.097& 0.045 &     0.804 & Galaxy & 155.0757 & 53.3210 & N/A \\ 
ZTF19abkfshj & 0.081  &    0.209     & 0.046& 0.663 & No Host & N/A & N/A & SLSN-I\\ 
ZTF18aawpuqx & 0.061 & 0.610 & 0.174 & 0.153 & Galaxy & 183.2727 &	35.5905 & N/A\\ 
ZTF21aakvrwp & 0.475& 0.187 &     0.319& 0.018 & Galaxy & 236.3290 &	31.6679 & N/A
\enddata
\caption{Sample of the entire catalog of 12,993 classified light curves and candidate host galaxies.}\label{chartable}

\end{deluxetable*}

\subsection{Wavelet Transformation and Other Features}

To model our supernova light curves we use a stationary wavelet transform \citep{lochner_photometric_2016}. We used a sampling of 100 on the GPR curve and a two-level wavelet transform, resulting in 800 coefficients per object. To deal with this high dimensionality, we use Principal Component Analysis (PCA), which reduced our dimensionality from 800 dimensions to 10. Along with these 10 wavelet features we calculate the area beneath the light curve, multi-band variance, duration, and range of the light curve.

\subsection{Host Galaxy Association} \label{Host}
We associated supernovae with candidate host galaxies following GHOST \citep{gagliano_ghost_2021}.  We modified GHOST so that instead of eliminating any objects it classifies as stars, it determines whether or not an event is a star by applying the same host galaxy association but with tighter constraints on the criteria for a star to be the host. This allows us to easily filter out impurities in our final data set in \S \ref{10,000 Supernovae}. We estimate a host misassociation rate of 7.25\%; see \citet{gagliano_ghost_2021} for more details.

\subsection{Classifier} \label{sec:imbalance}
We used the balanced random forest classifier from the imbalanced learn software \citep{JMLR:v18:16-365}. We used the random search cv function from scikit-learn \citep{scikit-learn} to tune the model and decided on the following parameters: 2,000 estimators, no bootstrap, maximum features of 4, maximum depth of 45, sample with replacement, and a class weight of balanced subsample.

\section{Results} \label{sec:results}

\subsection{Performance} \label{sec:performance}

To measure the performance of our classifier we used stratified k-fold cross validation in scikit-learn \citep{scikit-learn}. We used the weighted F1-score, which takes into consideration both the precision and recall while also accounting for class imbalance, to evaluate our classifier.

Our classifier was able to achieve an overall accuracy of 80\% and a weighted F1 score of 0.82 according to 3 fold cross validation.  Because our training data are biased towards bright, low redshift objects, and many of the ZTF light curves being evaluated are only partially complete, our results are expected to generally be only representative of the best case scenario for our classifier. 

\subsection{Comparison to Previous Works}
The most direct comparison to our classifier is the transient classifier within ALeRCE \citep{S_nchez_S_ez_2021}. According to their confusion matrix, they achieved the following values of median accuracy with upper and lower limits representing the 95th and 5th percentile, respectively: 76\%$\pm^7_6$ of Type Ia, 53\%$\pm^7_9$ of Type II, 50\%$\pm^{17}_6$ of Type Ib/c, and 100\%$\pm^0_{26}$ of SLSN. Our classifier achieved the following accuracy: 85\% of Type Ia, 60\% of Type II, 69\% of Type Ib/c, and 77\% of SLSN. Our classifier achieved better performance overall, while only using 14 features versus their 152 features. However, it should be noted that our classifier used a larger training set, which could be a source of difference in capability.   

\subsection{Catalog of Photometrically Classified Supernovae} \label{10,000 Supernovae}
Table \ref{chartable} shows results from our catalog of 12,993 objects that have been photometrically classified using our classifier (including our training set).\footnote{See  \citet{Zenodo} for complete listing and associated products.} Light curves were first obtained by querying the ALeRCE database \citep{S_nchez_S_ez_2021} for events that had a minimum of 10 alerts and a minimum of 0.7 probability of being a transient according to their classifier. These parameters were chosen based on our own experimentation. We found that 10 alerts provided, on average, sufficient coverage for our classifier to be useful, while still including a large portion of events. The minimum of 0.7 probability was chosen because we found that it provided an optimal balance between supernova purity and the overall number of light curves gathered.  Additionally, we used GHOST for star removal. When comparing the distribution of classes outside our training set we found it to consist of: 57.6\% Type Ia, 17.9\% Type II, 18.6\% Type Ib/c, and 6\% SLSN. To determine if this distribution is representative of the actual discovery rate of supernovae we compare it to the distribution reported by BTS \citep{Perley_2020} which yields: 72.7\% Type~Ia, 19.9\% Type~II, 5.9\% Type Ib/c, and 1.4\% SLSN. This indicates that our classifier overestimates the number of Type Ib/c while underestimating the number of Type Ia.

\section{Conclusion} \label{sec:conclusion}

We have presented a simple, yet effective, classification model that can achieve 80\% accuracy distinguishing SN Ia, II, Ib/c, and SLSN events in ZTF light curves. We have also provided a large set of 12,993, ZTF light curves that have been classified using only photometry and associated with candidate host galaxies. These products can be used to build upon current classification techniques, as well as enable population-scale studies of supernova explosions. 

\begin{acknowledgments}
D.~M.\ acknowledges NSF support from grants PHY-1914448 and AST-2037297. 
\end{acknowledgments}

\bibliographystyle{aasjournal}
%%\bibliography{bibliography.bib}

\begin{thebibliography}{}
\expandafter\ifx\csname natexlab\endcsname\relax\def\natexlab#1{#1}\fi
\providecommand{\url}[1]{\href{#1}{#1}}
\providecommand{\dodoi}[1]{doi:~\href{http://doi.org/#1}{\nolinkurl{#1}}}
\providecommand{\doeprint}[1]{\href{http://ascl.net/#1}{\nolinkurl{http://ascl.net/#1}}}
\providecommand{\doarXiv}[1]{\href{https://arxiv.org/abs/#1}{\nolinkurl{https://arxiv.org/abs/#1}}}

\bibitem[{{Ambikasaran} {et~al.}(2015){Ambikasaran}, {Foreman-Mackey},
  {Greengard}, {Hogg}, \& {O'Neil}}]{hodlr}
{Ambikasaran}, S., {Foreman-Mackey}, D., {Greengard}, L., {Hogg}, D.~W., \&
  {O'Neil}, M. 2015, IEEE Transactions on Pattern Analysis and Machine
  Intelligence, 38, 252, \dodoi{10.1109/TPAMI.2015.2448083}

\bibitem[{{Bellm} {et~al.}(2019){Bellm}, {Kulkarni}, {Graham}, {Dekany},
  {Smith}, {Riddle}, {Masci}, {Helou}, {Prince}, {Adams}, {Barbarino},
  {Barlow}, {Bauer}, {Beck}, {Belicki}, {Biswas}, {Blagorodnova}, {Bodewits},
  {Bolin}, {Brinnel}, {Brooke}, {Bue}, {Bulla}, {Burruss}, {Cenko}, {Chang},
  {Connolly}, {Coughlin}, {Cromer}, {Cunningham}, {De}, {Delacroix}, {Desai},
  {Duev}, {Eadie}, {Farnham}, {Feeney}, {Feindt}, {Flynn}, {Franckowiak},
  {Frederick}, {Fremling}, {Gal-Yam}, {Gezari}, {Giomi}, {Goldstein},
  {Golkhou}, {Goobar}, {Groom}, {Hacopians}, {Hale}, {Henning}, {Ho}, {Hover},
  {Howell}, {Hung}, {Huppenkothen}, {Imel}, {Ip}, {Ivezi{\'c}}, {Jackson},
  {Jones}, {Juric}, {Kasliwal}, {Kaspi}, {Kaye}, {Kelley}, {Kowalski},
  {Kramer}, {Kupfer}, {Landry}, {Laher}, {Lee}, {Lin}, {Lin}, {Lunnan},
  {Giomi}, {Mahabal}, {Mao}, {Miller}, {Monkewitz}, {Murphy}, {Ngeow},
  {Nordin}, {Nugent}, {Ofek}, {Patterson}, {Penprase}, {Porter}, {Rauch},
  {Rebbapragada}, {Reiley}, {Rigault}, {Rodriguez}, {van Roestel}, {Rusholme},
  {van Santen}, {Schulze}, {Shupe}, {Singer}, {Soumagnac}, {Stein}, {Surace},
  {Sollerman}, {Szkody}, {Taddia}, {Terek}, {Van Sistine}, {van Velzen},
  {Vestrand}, {Walters}, {Ward}, {Ye}, {Yu}, {Yan}, \& {Zolkower}}]{Bellm_2018}
{Bellm}, E.~C., {Kulkarni}, S.~R., {Graham}, M.~J., {et~al.} 2019, \pasp, 131,
  018002, \dodoi{10.1088/1538-3873/aaecbe}

\bibitem[{{Boone}(2019)}]{boone_avocado_2019}
{Boone}, K. 2019, \aj, 158, 257, \dodoi{10.3847/1538-3881/ab5182}

\bibitem[{{Gagliano} {et~al.}(2021){Gagliano}, {Narayan}, {Engel}, {Carrasco
  Kind}, \& {LSST Dark Energy Science Collaboration}}]{gagliano_ghost_2021}
{Gagliano}, A., {Narayan}, G., {Engel}, A., {Carrasco Kind}, M., \& {LSST Dark
  Energy Science Collaboration}. 2021, \apj, 908, 170,
  \dodoi{10.3847/1538-4357/abd02b}

\bibitem[{{Garretson} {et~al.}(2021){Garretson}, {Milisavljevic}, {Reynolds},
  {Weil}, {Subrayan}, {Banovetz}, \& {Lee}}]{Zenodo}
{Garretson}, B., {Milisavljevic}, D., {Reynolds}, J., {et~al.} 2021,
  \dodoi{10.5281/zenodo.5735569}

\bibitem[{{Ivezi{\'c}} {et~al.}(2019){Ivezi{\'c}}, {Kahn}, {Tyson}, {Abel},
  {Acosta}, {Allsman}, {Alonso}, {AlSayyad}, {Anderson}, {Andrew}, {Angel},
  {Angeli}, {Ansari}, {Antilogus}, {Araujo}, {Armstrong}, {Arndt}, {Astier},
  {Aubourg}, {Auza}, {Axelrod}, {Bard}, {Barr}, {Barrau}, {Bartlett}, {Bauer},
  {Bauman}, {Baumont}, {Bechtol}, {Bechtol}, {Becker}, {Becla}, {Beldica},
  {Bellavia}, {Bianco}, {Biswas}, {Blanc}, {Blazek}, {Blandford}, {Bloom},
  {Bogart}, {Bond}, {Booth}, {Borgland}, {Borne}, {Bosch}, {Boutigny},
  {Brackett}, {Bradshaw}, {Brandt}, {Brown}, {Bullock}, {Burchat}, {Burke},
  {Cagnoli}, {Calabrese}, {Callahan}, {Callen}, {Carlin}, {Carlson},
  {Chandrasekharan}, {Charles-Emerson}, {Chesley}, {Cheu}, {Chiang}, {Chiang},
  {Chirino}, {Chow}, {Ciardi}, {Claver}, {Cohen-Tanugi}, {Cockrum}, {Coles},
  {Connolly}, {Cook}, {Cooray}, {Covey}, {Cribbs}, {Cui}, {Cutri}, {Daly},
  {Daniel}, {Daruich}, {Daubard}, {Daues}, {Dawson}, {Delgado}, {Dellapenna},
  {de Peyster}, {de Val-Borro}, {Digel}, {Doherty}, {Dubois},
  {Dubois-Felsmann}, {Durech}, {Economou}, {Eifler}, {Eracleous}, {Emmons},
  {Fausti Neto}, {Ferguson}, {Figueroa}, {Fisher-Levine}, {Focke}, {Foss},
  {Frank}, {Freemon}, {Gangler}, {Gawiser}, {Geary}, {Gee}, {Geha}, {Gessner},
  {Gibson}, {Gilmore}, {Glanzman}, {Glick}, {Goldina}, {Goldstein}, {Goodenow},
  {Graham}, {Gressler}, {Gris}, {Guy}, {Guyonnet}, {Haller}, {Harris},
  {Hascall}, {Haupt}, {Hernandez}, {Herrmann}, {Hileman}, {Hoblitt}, {Hodgson},
  {Hogan}, {Howard}, {Huang}, {Huffer}, {Ingraham}, {Innes}, {Jacoby}, {Jain},
  {Jammes}, {Jee}, {Jenness}, {Jernigan}, {Jevremovi{\'c}}, {Johns}, {Johnson},
  {Johnson}, {Jones}, {Juramy-Gilles}, {Juri{\'c}}, {Kalirai}, {Kallivayalil},
  {Kalmbach}, {Kantor}, {Karst}, {Kasliwal}, {Kelly}, {Kessler}, {Kinnison},
  {Kirkby}, {Knox}, {Kotov}, {Krabbendam}, {Krughoff}, {Kub{\'a}nek},
  {Kuczewski}, {Kulkarni}, {Ku}, {Kurita}, {Lage}, {Lambert}, {Lange},
  {Langton}, {Le Guillou}, {Levine}, {Liang}, {Lim}, {Lintott}, {Long},
  {Lopez}, {Lotz}, {Lupton}, {Lust}, {MacArthur}, {Mahabal}, {Mandelbaum},
  {Markiewicz}, {Marsh}, {Marshall}, {Marshall}, {May}, {McKercher}, {McQueen},
  {Meyers}, {Migliore}, {Miller}, {Mills}, {Miraval}, {Moeyens}, {Moolekamp},
  {Monet}, {Moniez}, {Monkewitz}, {Montgomery}, {Morrison}, {Mueller},
  {Muller}, {Mu{\~n}oz Arancibia}, {Neill}, {Newbry}, {Nief}, {Nomerotski},
  {Nordby}, {O'Connor}, {Oliver}, {Olivier}, {Olsen}, {O'Mullane}, {Ortiz},
  {Osier}, {Owen}, {Pain}, {Palecek}, {Parejko}, {Parsons}, {Pease},
  {Peterson}, {Peterson}, {Petravick}, {Libby Petrick}, {Petry},
  {Pierfederici}, {Pietrowicz}, {Pike}, {Pinto}, {Plante}, {Plate}, {Plutchak},
  {Price}, {Prouza}, {Radeka}, {Rajagopal}, {Rasmussen}, {Regnault}, {Reil},
  {Reiss}, {Reuter}, {Ridgway}, {Riot}, {Ritz}, {Robinson}, {Roby}, {Roodman},
  {Rosing}, {Roucelle}, {Rumore}, {Russo}, {Saha}, {Sassolas}, {Schalk},
  {Schellart}, {Schindler}, {Schmidt}, {Schneider}, {Schneider}, {Schoening},
  {Schumacher}, {Schwamb}, {Sebag}, {Selvy}, {Sembroski}, {Seppala}, {Serio},
  {Serrano}, {Shaw}, {Shipsey}, {Sick}, {Silvestri}, {Slater}, {Smith},
  {Smith}, {Sobhani}, {Soldahl}, {Storrie-Lombardi}, {Stover}, {Strauss},
  {Street}, {Stubbs}, {Sullivan}, {Sweeney}, {Swinbank}, {Szalay}, {Takacs},
  {Tether}, {Thaler}, {Thayer}, {Thomas}, {Thornton}, {Thukral}, {Tice},
  {Trilling}, {Turri}, {Van Berg}, {Vanden Berk}, {Vetter}, {Virieux},
  {Vucina}, {Wahl}, {Walkowicz}, {Walsh}, {Walter}, {Wang}, {Wang}, {Warner},
  {Wiecha}, {Willman}, {Winters}, {Wittman}, {Wolff}, {Wood-Vasey}, {Wu},
  {Xin}, {Yoachim}, \& {Zhan}}]{Ivezi__2019}
{Ivezi{\'c}}, {\v{Z}}., {Kahn}, S.~M., {Tyson}, J.~A., {et~al.} 2019, \apj,
  873, 111, \dodoi{10.3847/1538-4357/ab042c}

\bibitem[{{Lemaitre} {et~al.}(2016){Lemaitre}, {Nogueira}, \&
  {Aridas}}]{JMLR:v18:16-365}
{Lemaitre}, G., {Nogueira}, F., \& {Aridas}, C.~K. 2016, arXiv e-prints,
  arXiv:1609.06570.
\newblock \doarXiv{1609.06570}

\bibitem[{{Lochner} {et~al.}(2016){Lochner}, {McEwen}, {Peiris}, {Lahav}, \&
  {Winter}}]{lochner_photometric_2016}
{Lochner}, M., {McEwen}, J.~D., {Peiris}, H.~V., {Lahav}, O., \& {Winter},
  M.~K. 2016, \apjs, 225, 31, \dodoi{10.3847/0067-0049/225/2/31}

\bibitem[{{Pedregosa} {et~al.}(2012){Pedregosa}, {Varoquaux}, {Gramfort},
  {Michel}, {Thirion}, {Grisel}, {Blondel}, {M{\"u}ller}, {Nothman}, {Louppe},
  {Prettenhofer}, {Weiss}, {Dubourg}, {Vanderplas}, {Passos}, {Cournapeau},
  {Brucher}, {Perrot}, \& {Duchesnay}}]{scikit-learn}
{Pedregosa}, F., {Varoquaux}, G., {Gramfort}, A., {et~al.} 2012, arXiv
  e-prints, arXiv:1201.0490.
\newblock \doarXiv{1201.0490}

\bibitem[{{Perley} {et~al.}(2020){Perley}, {Fremling}, {Sollerman}, {Miller},
  {Dahiwale}, {Sharma}, {Bellm}, {Biswas}, {Brink}, {Bruch}, {De}, {Dekany},
  {Drake}, {Duev}, {Filippenko}, {Gal-Yam}, {Goobar}, {Graham}, {Graham}, {Ho},
  {Irani}, {Kasliwal}, {Kim}, {Kulkarni}, {Mahabal}, {Masci}, {Modak}, {Neill},
  {Nordin}, {Riddle}, {Soumagnac}, {Strotjohann}, {Schulze}, {Taggart},
  {Tzanidakis}, {Walters}, \& {Yan}}]{Perley_2020}
{Perley}, D.~A., {Fremling}, C., {Sollerman}, J., {et~al.} 2020, \apj, 904, 35,
  \dodoi{10.3847/1538-4357/abbd98}

\bibitem[{{S{\'a}nchez-S{\'a}ez} {et~al.}(2021){S{\'a}nchez-S{\'a}ez}, {Reyes},
  {Valenzuela}, {F{\"o}rster}, {Eyheramendy}, {Elorrieta}, {Bauer},
  {Cabrera-Vives}, {Est{\'e}vez}, {Catelan}, {Pignata}, {Huijse}, {De Cicco},
  {Ar{\'e}valo}, {Carrasco-Davis}, {Abril}, {Kurtev}, {Borissova}, {Arredondo},
  {Castillo-Navarrete}, {Rodriguez}, {Ruz-Mieres}, {Moya},
  {Sabatini-Gacit{\'u}a}, {Sep{\'u}lveda-Cobo}, \&
  {Camacho-I{\~n}iguez}}]{S_nchez_S_ez_2021}
{S{\'a}nchez-S{\'a}ez}, P., {Reyes}, I., {Valenzuela}, C., {et~al.} 2021, \aj,
  161, 141, \dodoi{10.3847/1538-3881/abd5c1}

\bibitem[{{Schlegel} {et~al.}(1998){Schlegel}, {Finkbeiner}, \&
  {Davis}}]{1998ApJ...500..525S}
{Schlegel}, D.~J., {Finkbeiner}, D.~P., \& {Davis}, M. 1998, \apj, 500, 525,
  \dodoi{10.1086/305772}

\bibitem[{{Sravan} {et~al.}(2020){Sravan}, {Milisavljevic}, {Reynolds},
  {Lentner}, \& {Linvill}}]{sravan_real-time_2020}
{Sravan}, N., {Milisavljevic}, D., {Reynolds}, J.~M., {Lentner}, G., \&
  {Linvill}, M. 2020, \apj, 893, 127, \dodoi{10.3847/1538-4357/ab8128}

\bibitem[{{Villar} {et~al.}(2020){Villar}, {Hosseinzadeh}, {Berger},
  {Ntampaka}, {Jones}, {Challis}, {Chornock}, {Drout}, {Foley}, {Kirshner},
  {Lunnan}, {Margutti}, {Milisavljevic}, {Sanders}, {Pan}, {Rest}, {Scolnic},
  {Magnier}, {Metcalfe}, {Wainscoat}, \& {Waters}}]{villar_superraenn_2020}
{Villar}, V.~A., {Hosseinzadeh}, G., {Berger}, E., {et~al.} 2020, \apj, 905,
  94, \dodoi{10.3847/1538-4357/abc6fd}

\end{thebibliography}

\end{document}